\documentclass[runningheads]{llncs}
\usepackage[compress, sort]{cite} 
\usepackage{graphicx}
\usepackage[title]{appendix} 
\usepackage{amsmath} 
\usepackage{cases} 
\usepackage{xspace}
\usepackage{bm}
\usepackage{multirow}
\usepackage{booktabs}
\usepackage{amssymb}
\usepackage{wrapfig}
\usepackage{arydshln}
\newif\ifanonymous
\newcommand{\vecv}{\boldsymbol{v}}
\newcommand{\vecp}{\boldsymbol{p}}
\newcommand{\Leul}{\ensuremath{\mathcal{L}_\mathrm{eul}}\xspace}
\newcommand{\Llag}{\ensuremath{\mathcal{L}_\mathrm{lag}}\xspace}
\newcommand{\Lsim}{\ensuremath{\mathcal{L}_\mathrm{sim}}\xspace}
\newcommand{\Lsm}{\ensuremath{\mathcal{L}_\mathrm{smooth}}\xspace}
\newcommand{\Linc}{\ensuremath{\mathcal{L}_\mathrm{inc}}\xspace}
\newcommand{\para}[1]{\noindent\textbf{#1}\quad}
\newcommand{\pp}{\scriptsize **}
\newcommand{\p}{\scriptsize *}

\begin{document}

\title{MomentaMorph: Unsupervised Spatial-Temporal Registration with Momenta, Shooting, and Correction}
\titlerunning{MomentaMorph}

\ifanonymous
	\author{\,}
	\authorrunning{anonymous submission}
	\institute{\,}
\else
\author{Zhangxing Bian\inst{1}\orcidID{0000-0003-3603-0650} \and Shuwen Wei\orcidID{0000-0001-8679-9615}\inst{1} \and \\ Yihao Liu\inst{1} \and Junyu Chen\inst{2} \and  Jiachen Zhuo\inst{4} \and Fangxu Xing\inst{3} \and Jonghye Woo\inst{3} \and  Aaron Carass\inst{1}\orcidID{0000-0003-4939-5085} \and Jerry~L.~Prince\inst{1}\orcidID{0000-0002-6553-0876}
 }
\authorrunning{Z. Bian et al.}
%
\institute{Johns Hopkins University, Baltimore MD, USA \\
	\email{\{zbian4,swei14,yliu236,aaron\_carass,prince\}@jhu.edu}\\ \and
	Johns Hopkins School of Medicine, Baltimore MD, USA \\
	\email{jchen245@jhmi.edu}\\ \and
	Massachusetts General Hospital and Harvard Medical School, Boston MA, USA
	\email{\{fxing1,jwoo\}@mgh.harvard.edu}\\ \and
University of Maryland School of Medicine, Baltimore MD, USA \\
\email{jzhuo@umm.edu}
}
\fi
\maketitle              
\ifanonymous
	\vspace{-5em}
\fi
\begin{abstract}
Tagged magnetic resonance imaging~(tMRI) has been employed for decades to measure the motion of tissue undergoing deformation. 
However, registration-based motion estimation from tMRI is difficult due to the periodic patterns in these images, particularly when the motion is large.
With a larger motion the registration approach gets trapped in a local optima, leading to motion estimation errors. 
We introduce a novel ``momenta, shooting, and correction" framework for Lagrangian motion estimation in the presence of repetitive patterns and large motion.
This framework, grounded in Lie algebra and Lie group principles, accumulates momenta in the tangent vector space and employs exponential mapping in the diffeomorphic space for rapid approximation towards true optima, circumventing local optima.
A subsequent correction step ensures convergence to true optima.
The results on  a 2D synthetic dataset and a real 3D tMRI dataset demonstrate our method's efficiency in estimating accurate, dense, and diffeomorphic 2D/3D motion fields amidst large motion and repetitive patterns. 

\end{abstract}
\section{Introduction}

\begin{figure}[tb]
	\includegraphics[width = 0.99\linewidth]{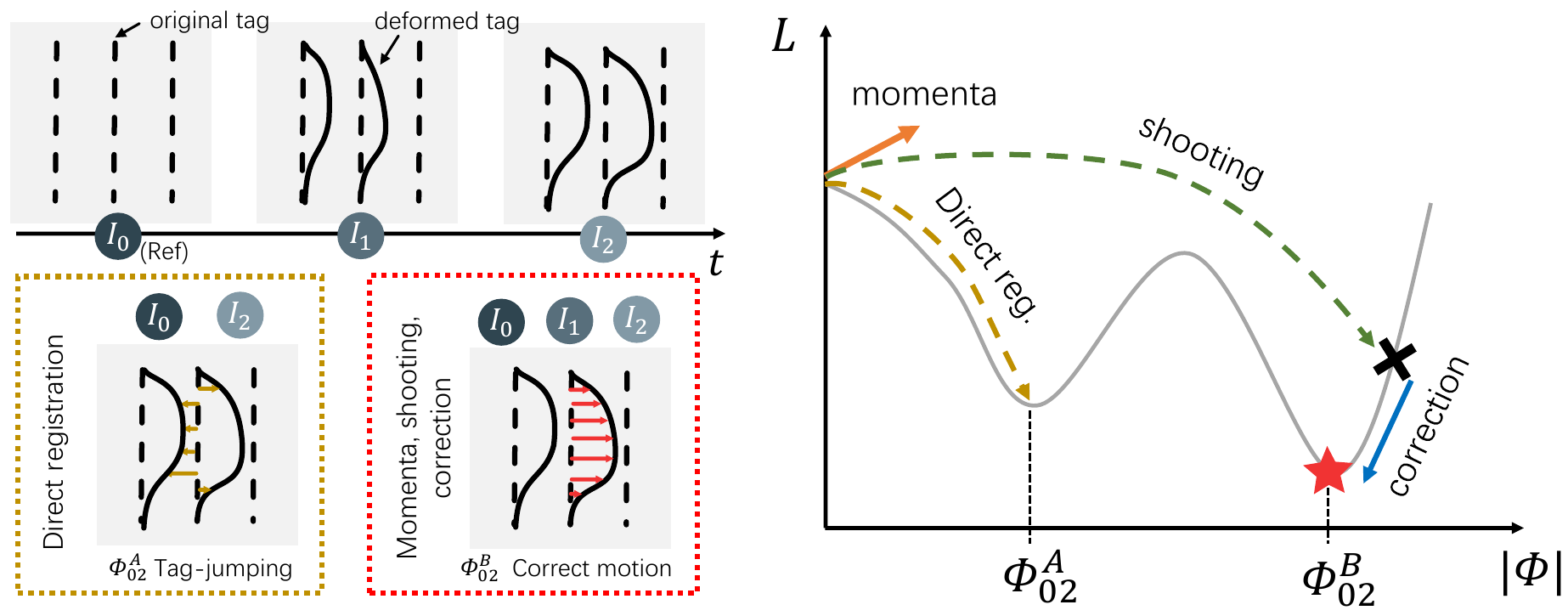}
	\caption{Conceptual illustration of tag-jumping in unsupervised registration. The left panel demonstrates two strategies for estimating motion between the reference timeframe $I_0$ and the moving timeframe $I_2$: Method A~($\Phi^{A}_{02}$) directly registers $I_0$ and $I_2$, while Method B~($\Phi^{B}_{02}$) utilizes the temporal information from $I_0$ \textit{thru} $I_2$ for registration. 
    The right panel visualizes the loss landscape~($L$ on $y$-axis) in optimization, highlighting local optima caused by image periodicity in $\Phi$ space.}
	\label{fig:intro_tagjumping}
\end{figure}

Tagging magnetic resonance imaging~(tMRI)~\cite{axel1989heart, axel1989mr} is an important imaging technique, enabling precise measurements and visualizations of tissue movement and deformation.
However, the tagged periodic patterns present considerable challenges in estimating Lagrangian motion.
Lagrangian motion traces tissues from their resting state and is especially suited for strain computation~\cite{ibrahim2011myocardial, knutsen2014improved, gomez2020analysis, shao2023analysis}.
Existing approaches typically employ raw tagged~\cite{ye2021deeptag} or harmonic phase~\cite{osman1999cardiac, PVIRA, bian2023drimet} images which \textit{all} have periodic patterns as inputs to unsupervised registration algorithms~\cite{ilogdemons2011, voxelmorph2019}, to estimate the Lagrangian motion.
However, when the deformation between the registration pair exceeds half of the tagging period, these methods risk falling into local optima, leading to substantial errors---a phenomenon known as ``tag-jumping"---as illustrated in Fig.~\ref{fig:intro_tagjumping}.

Several strategies have been proposed to address this issue. For instance, Shortest Path Refinement~\cite{liu2010shortest} incorporates a static region assumption into a region growing algorithm to correct the motion estimation.
This approach requires strong assumptions and human intervention. 
Phase unwrapping~\cite{spoorthi2018phasenet, wang2022deep}, another strategy, unwraps the periodic patterns before registration.
Yet, it is often susceptible to image noise and artifacts~\cite{jenkinson2003fast, wang2022deep}. 
Deep learning-based pattern unwrapping methods, though demonstrating better accuracy, are mostly trained on synthetic images, raising questions about their generalizability to real data. While many methods primarily utilize pair frames, it is possible to leverage the entire sequence. Yu~\emph{et al.}~\cite{NewStartPoint2023} utilize traditional iterative registration~\cite{PVIRA}, proposing to initialize the registration with composition of motion between every successive pair of frames. However, this approach incurs high computational costs due to its iterative nature and on-the-fly optimization.

Unsupervised deep learning-based registration methods have recently shown promise, offering good registration accuracy, fast inference speed~\cite{dalca2018unsupervised,  hering2021cnn, im2grid2022, ye2021deeptag, bian2023drimet, chen2023survey}.
Ye \emph{et al.}~\cite{ye2021deeptag} propose to estimate Lagrangian motion from raw tMRI by composing small Eulerian motions.
However, this can lead to drifting issues, where small errors in each step accumulate in the composed Lagrangian motion, leading to inferior accuracy. 
DRIMET~\cite{bian2023drimet} proposed a penalty on learning the incompressible motion in tMRI, yet like many unsupervised registration algorithms that focus on registering frame pairs, it struggles with tag-jumping. 

\begin{wrapfigure}{l}{0.45\textwidth} 
	\centering
	\includegraphics[width= \linewidth]{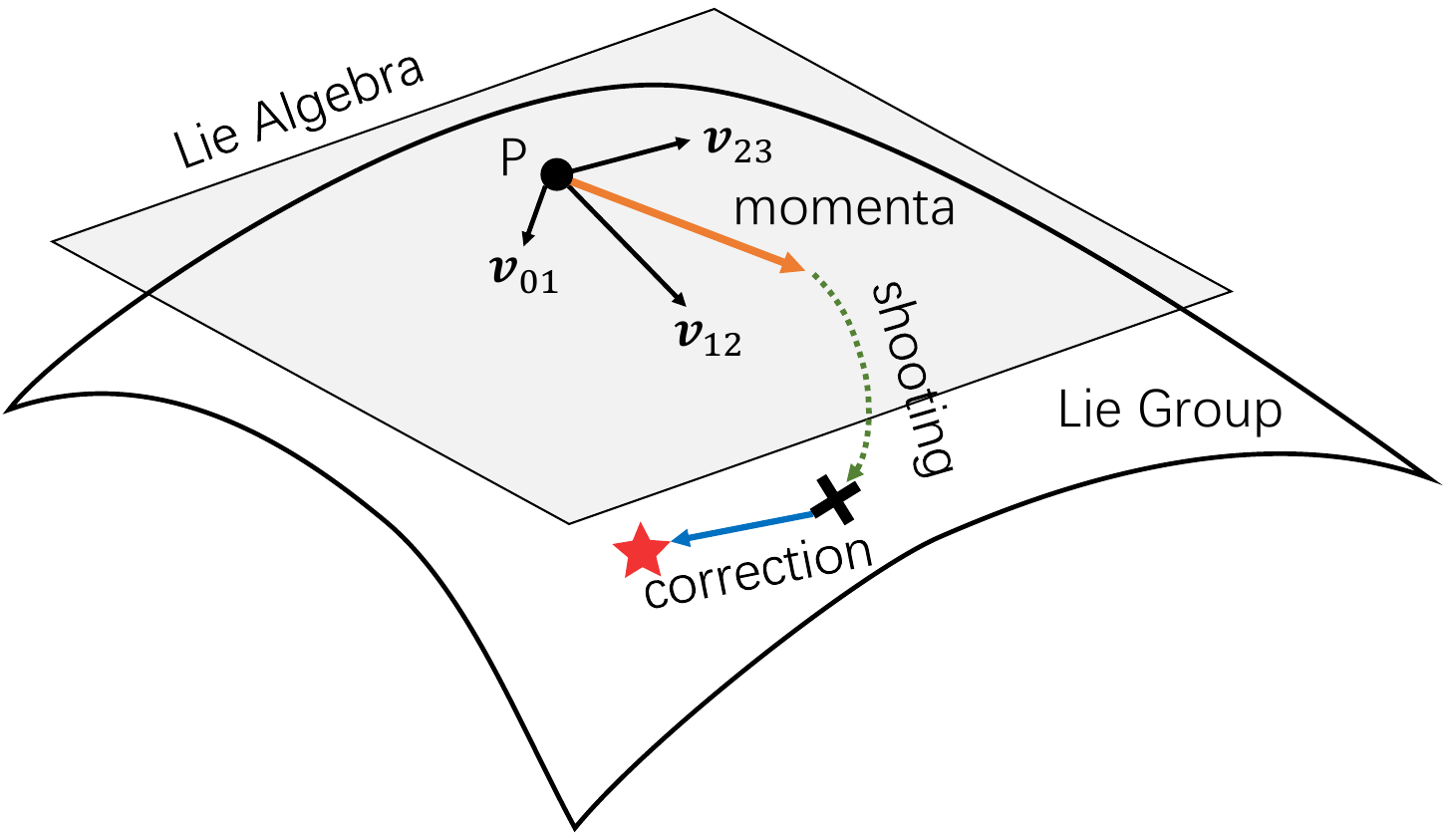}
	\vspace{-1em}
	\caption{A high-level overview of  ``momenta, shooting, and correction".}
	\label{fig:high-level}
	\vspace{-2.5em}
\end{wrapfigure}

Inspired by these previous works and the principle of ``fast-slow optimization'' in look-ahead optimizers~\cite{zhang2019lookahead}, we introduce ``MomentaMorph: momenta, shooting, and correction'', a novel framework for estimating Lagrangian motion in the presence of repetitive patterns and large motions. 
As shown in Fig.~\ref{fig:high-level}, this framework operates under Lie algebra and Lie group principles, where momenta accumulate in the tangent vector space~(Lie algebra) and shooting occurs in the diffeomorphic space (Lie group) through exponential mapping, serving as the ``fast'' step in ``fast-slow'' optimization.
This process ``shoots'' towards an approximation near the true optima, avoiding the local optima presented by large motions in repetitive patterns.
A subsequent correction step ensures convergence to the true optima, thereby enhancing accuracy and acting as the ``slow'' step.

While MomentaMorph shares similarity in nomenclature with LDDMM~\cite{vialard2012diffeomorphic}, there are differences.
LDDMM estimates an initial momentum field that, when ``shot'' forward via a geodesic path within the context of a diffeomorphism, deforms one image to align with another.
However, it does not exploit temporal information from multiple frames and carries the risk of falling into local optima during the ``shooting" process.
It also requires a precise initial estimate of the momentum and can be computationally intensive due to the necessity of solving a partial differential equation at each iteration.
In contrast, MomentaMorph exploits temporal information from multiple frames and is an amortized learning-based approach, enabling rapid inference once trained. 
Our context also differs from natural video tracking~\cite{araslanov2021dense, bian2022learning} or optical flow~\cite{horn1981determining, jonschkowski2020matters, stone2021smurf, bian2022learning} in computer vision, which typically have less emphasis on incompressibility, diffeomorphisms, or Lagrangian motion.

Our contribution is that we introduce a novel framework MomentaMorph for motion tracking using unsupervised registration techniques.
MomentaMorph can estimate accurate, diffeomorphic, and incompressible Lagrangian motion. 
MomentaMorph is validated using both a 2D synthetic motion dataset and a real 3D tMRI dataset capturing tongue motion, demonstrating its broad applicability. 
Broadly speaking, this work also offers a promising solution for harnessing the temporal information inherent in biological dynamic imaging modalities.

\section{Method}
Drawing on previous research~\cite{christensen1996deformable, beg2005computing, arsigny2006log, dalca2018unsupervised}, we define the deformation field via an ordinary differential equation~(ODE) $\frac{\partial \phi^{(\tau)}}{\partial \tau}=\boldsymbol{v} \circ \phi^{(\tau)}$, where $\phi^{(0)}$ represents the identity transformation and $\vecv$ is a stationary velocity field~(SVF).
One can integrate $\boldsymbol{v}$ over $\tau=[0,1]$ to obtain the ﬁnal registration ﬁeld $\phi^{(1)}$.
In the context of group theory, $\vecv$ belongs to the Lie algebra and yields a member of the Lie group through the exponential map $\phi = \exp(\boldsymbol{v})$.

\begin{figure}[!tb]
	\centering
	\includegraphics[width= 0.9\linewidth]{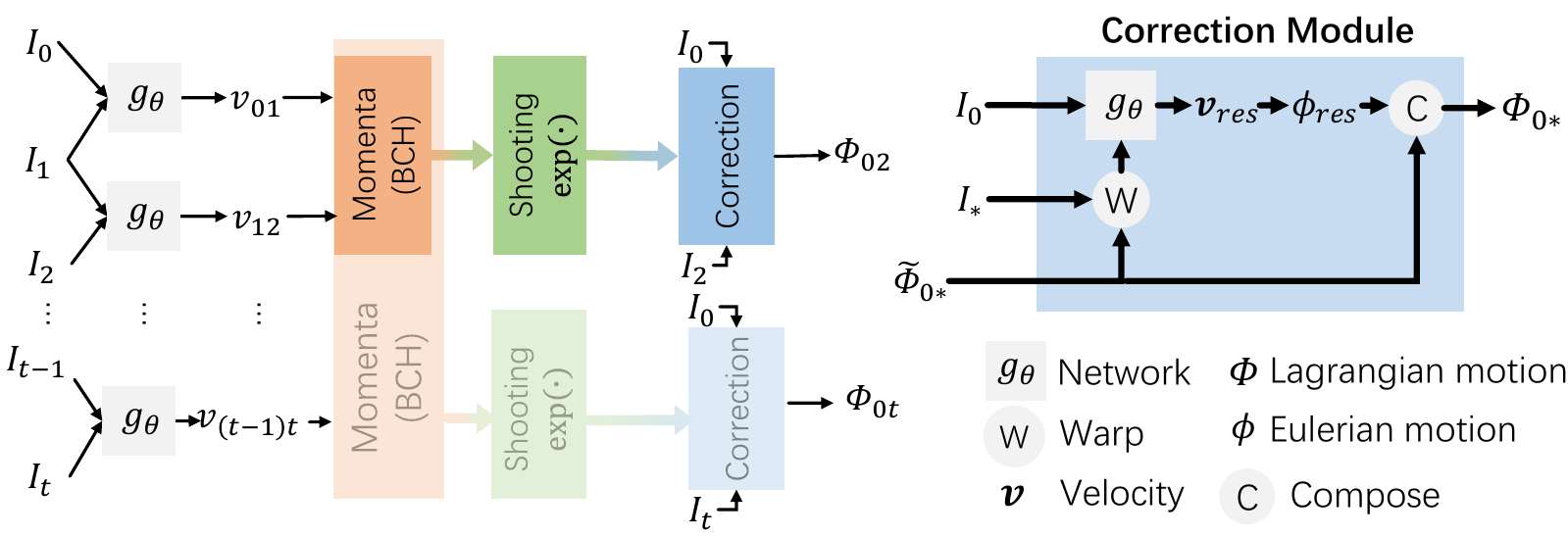}
	\caption{The architecture of MomentaMorph. The input is a sequence of images $\{I_0, I_1, I_2, \dots\}$, and the output is the Lagrangian motion between the reference frame and others. A detailed view of the correction module is shown on the right.}
	\label{fig:net_pipeline}
\end{figure}

Consider $\{I_0, I_1, I_2, \dots \}$ as a sequence of 3D images, such as a tMRI sequence. By convention, $I_0$ is selected as our reference frame.
The motion field between consecutive frames $t$ and $t + 1$ is referred to as Eulerian motion, denoted as $\phi_{t(t+1)}$,  
while the motion field between the reference frame $I_0$ and any other frame $I_t$ is referred to as Lagrangian motion, denoted as $\Phi_{0t}$.
We propose an unsupervised framework to estimate the 3D dense diffeomorphic Lagrangian motion fields given the sequence $\{I_0, I_1, I_2, \dots\}$.

For frames $I_t$ and $I_{t+1}$, $\vecv_{t(t+1)} = g_\theta(I_t,I_{t+1})$ estimates the SVF, where $g$ is a convolutional neural network with learnable parameters $\theta$.
We assume that there is no tag-jumping problems between successive frames when imaging settings are appropriately configured. 
To obtain Lagrangian motion, one might consider composing the Eulerian motions.
However, such an approach can lead to drifting issues, resulting in inaccurate motion.
Instead, we introduce the momenta, shooting, and correction framework, as shown in Fig.~\ref{fig:net_pipeline}.

\subsection{Momenta, shooting, correction}

\textbf{Momenta} The Lagrangian motion can be expressed in terms of velocity fields, for $t \geq 1 $, as follows:
\begin{subequations}
\begin{align}
	\log \left(\Phi_{0 t}\right)=\vecv_{0 t} & =\log \left(\phi_{(t-1) t} \circ \Phi_{0(t-1)}\right) \label{eq:1a}\\
	& =\log \left(\exp \left(\vecv_{(t-1) t}\right) \circ \exp \left(\vecv_{0(t-1)}\right)\right) \label{eq:1b}\\
	& =\vecv_{(t-1) t}+\vecv_{0(t-1)}+ \frac{1}{2} \left[\vecv_{(t-1) t}, \vecv_{0(t-1)}\right] + \cdots \label{eq:1c}\\
	 \text{ with } \log \left(\Phi_{00}\right)&=\vecv_{00} =\bm{0} .   \notag 
\end{align}
\end{subequations}

\noindent Equations~(\ref{eq:1a}-\ref{eq:1b}) represents a recursive formula, valid for $t \geq 1$, indicating that each $\Phi_{0t}$ can be expressed as a composition involving the preceding term $\Phi_{0(t-1)}$ and a new term $\phi_{(t-1)t}$, and thus can be expressed with elements in Lie algebras.
Equation~(\ref{eq:1c}) applies the Baker-Campbell-Hausdorff~(BCH) formula~\cite{baker1897abel, campbell1926course, hall2013lie}, which establishes a connection between the product of exponentials and associative algebras. 
Specifically, $\circ$ represents a composition map associated with the Lie group.
The operator $[\cdot,\cdot]$ denotes the Lie bracket.
The dots indicate higher-order terms involving repeated Lie brackets, which are omitted for simplicity. 

Although the BCH formula theoretically converges only when the two velocity fields are sufficiently small, it has been observed to perform well in broader situations. 
Using only the first few terms in the series defined in Eq.~(\ref{eq:1c}), we form two approximations of momenta (for $t\ge 1$), as follows:
%
%
\begin{small}
	\begin{subequations}
		\begin{align}
			\text{[Mmorph-1]} \quad & \vecp^1_{0t} := \vecv_{(t-1)t} + \vecp^1_{0(t-1)} \label{eq:BCH1}\\
			\text{[Mmorph-2]}\quad & \vecp^2_{0t} :=   \vecv_{(t-1)t} + \vecp^2_{0(t-1)}  + \frac{1}{2}\,\left(J_{\vecp^2_{0(t-1)} }\vecv_{(t-1)t} - J_{\vecv_{(t-1)t}}\vecp^2_{0(t-1)}\right) \label{eq:BCH2} \\
			& \text{ with } \vecp^1_{00} = \vecp^2_{00} = \bm{0}, \nonumber
		\end{align}
	\end{subequations}
\end{small}

\noindent where $J_{\vecv}$ is the $3\times 3$ Jacobian matrix of vector field $\vecv$.

\noindent \textbf{Shooting} From either ({\ref{eq:BCH1}}) or ({\ref{eq:BCH2}}), we use the scaling and squaring~\cite{arsigny2006log} to compute the exponential mapping ${\widetilde\Phi_{0t}} = \exp(\vecp_{0t})$.
This shoots solutions to an approximate location near the true deformation on the manifold.
The approximation arise from two sources: the omitted high-order terms in Eq.~(\ref{eq:1c}), and the interpolation during the scaling and squaring on the discretized image grid.

\noindent \textbf{Correction}  We refine the Lagrangian estimate by first estimating the residual motion $\phi_\mathrm{res} = \exp(\vecv_\mathrm{res})$ where $\vecv_{\mathrm{res}} = g_\theta(I_0,I_t \circ {\widetilde\Phi_{0t}})$, and then composing with $\widetilde\Phi_{0t}$ to obtain the final result $\Phi_{0t} = { \widetilde\Phi_{0t} \circ  \phi_\mathrm{res} }$.

\subsection{Learning} 

Our overall training objective is $\mathcal{L} = \Leul + \gamma \, \Llag$ which involves both Eulerian motion, Eq.~(\ref{eq:Leul}), and Lagrangian motion, Eq.~(\ref{eq:Llag}), where
\begin{align}
\Leul &= \sum_{t=0}^{T} \Lsim(I_t, I_{t+1} \circ \phi_{t(t+1)}) + \alpha \Lsm(\phi_{t(t+1)}) + \beta \Linc(\phi_{t(t+1)}) \label{eq:Leul}\\
\Llag &= \sum_{t=1}^{T} \Lsim(I_0,I_t \circ \Phi_{0t})  + \alpha \Lsm(\Phi_{0t}) + \beta \Linc(\Phi_{0t}) \,. \label{eq:Llag}
\end{align}
In the above equations, $\gamma$, $\alpha$'s, and $\beta$'s are hyper-parameters, determined by grid search.
$\Lsim$, $\Lsm$, and $\Linc$ denote similarity loss, smoothness loss, and incompressibility loss, respectively. 
During training, we employ mean squared error~(MSE) as our similarity loss. 
We encourage the spatial smoothness of the displacement $\bm{u}$, with the smoothness loss defined as $\Lsm = \sum_{\bm{x}} | \nabla \bm{u}(\bm{x}) |^2$. 
Incompressibility is a crucial characteristic for image registration in moving biological tissues, such as myocardium muscles, the tongue, and the brain. 
The Jacobian determinant of a deformation, representing the ratio of volume change, is frequently utilized to quantify growth or shrinkage in biological tissue~\cite{rohlfing2003volume, bian2021deformable, burris2022vascular}.
We adopt the determinant-based penalty proposed by DRIMET~\cite{bian2023drimet} $\Linc = \sum_{\bm{x}} \left| \log \max \left( \left |J_\phi(\bm{x}) \right| ,\epsilon \right) \right| - \sum_{\bm{x}} \min \left( \left| J_\phi(\bm{x}) \right| , 0 \right)$, where $\left |J_\phi(\bm{x}) \right|$ is the Jacobian determinant of $\phi$ at $\bm{x}$. The $\Linc$ pushes the Jacobian determinant toward unity, encouraging incompressibility in the deformation field.

The training process is end-to-end, leveraging the differentiability of Eq.~(\ref{eq:BCH1}) and~(\ref{eq:BCH2}), exponential mapping, and composition with respect to the velocity field estimated by our network.
The  $\Leul$ guides the network to learn effective deformation fields for neighboring frames. 
Simultaneously, $\Llag$ encourages the network to predict a temporally-consistent Lagrangian motion field that can match the reference timeframe with any other timeframes, while promoting diffeomorphism and incompressible characteristics.

\section{Experiments}

\subsection{Synthetic Elastic 2D Data}

\para{Materials} We synthesized a dataset of 10,000 ``movies'', where each movie is composed of three sequential images with dimensions \(96 \times 96\). 
These images undergo elastic deformation. 
We ensure that the maximum deformation between consecutive frames does not exceed the theoretical critical point for tag-jumping~($P/2$). 
Those deformation fields are subsequently used to deform simulated sinusoidal images (both horizontal and vertical patterns), resulting in a time-series of deformed images.
The data samples were divided into training, validation, and test datasets in a 6:2:2 ratio.\footnote{See Appendix~\ref{app:exp} for more details on simulation and training.}

\begin{figure}[!tb]
	\centerline{\includegraphics[width = 0.99\linewidth]{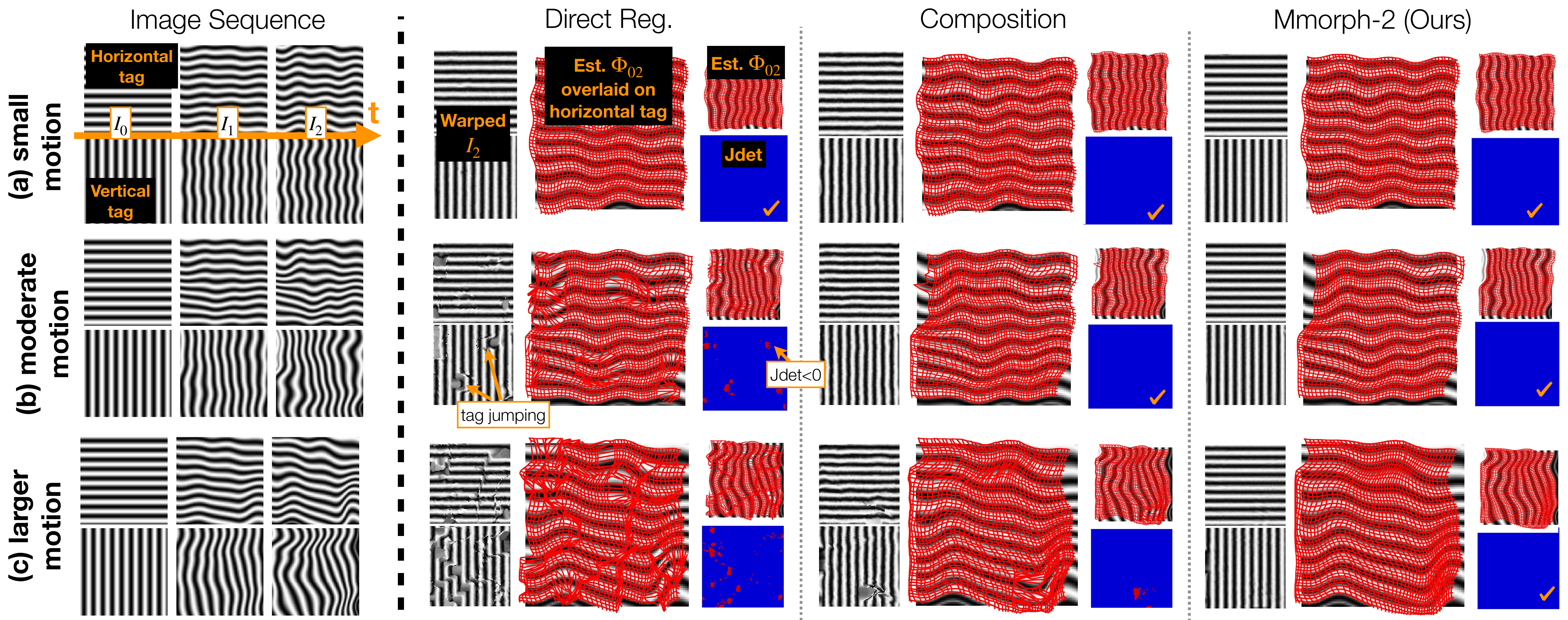}}
	\caption{Qualitative results on synthetic data. The first column displays the input image sequence, while the other columns correspond to the results of different approaches. Row~(a) to~(c) demonstrate an increasing amount of motion. The red areas on the Jacobian determinant map highlight the locations where foldings occur, and the symbol  $\checkmark$ represents a pure diffeomorphism. We have omitted the results of Mmorph-1 as they are visually similar to those of Mmorph-2.}
	\label{fig:Quali_2D}
\end{figure}

\begin{table}[!tb]
	\caption{Performance comparison on synthetic 2D data. Wilcoxon signed-rank tests was conducted between ``Mmorph-2" and the other methods. The p-values are adjusted by Bonferroni correction ($\times 3$). The symbol {\pp} indicates a statistically significant improvement ({\pp}: $p$-value $< 0.001$; {\p}: $p$-value $< 0.01$).  $\mathbf{T}$ indicates how many timeframes is used by each approach. ``Mmorph'' stands for MomentaMorph.}
	\label{tab:Quanti_2D}
	\centering
	\resizebox{0.99\textwidth}{!}{%
		\begin{tabular}{cccccccccccc}
			\toprule
			& & \multicolumn{3}{c}{\textbf{RMSE} $\downarrow$ } & \multicolumn{3}{c}{\textbf{EPE (pix)} $\downarrow$} & \multicolumn{2}{c}{\textbf{NegDet (\%)} $\downarrow$} & \textbf{Time(s) $\downarrow$} \\ 
			\cmidrule(r){3-5} \cmidrule(lr){6-8} \cmidrule(l){9-10} \cmidrule(l){11-11}
			& $\mathbf{T}$ & Mean \scriptsize$\pm$ Std & Median & $p$ & Mean \scriptsize $\pm$ Std & Median & $p$ & Mean \scriptsize $\pm$ Std & $p$ & Mean \\ 
			\midrule
			Direct Reg\cite{LKUnet2022}& 2 & 0.096 \scriptsize $\pm$ 0.053 & 0.098 & \pp &  2.491 $\pm$ \scriptsize 1.861 & 1.802 & \pp &  1.926 $\pm$ \scriptsize 1.842 & \pp & 0.04 \\
			Composition& 3 & 0.073 $\pm$ \scriptsize 0.025 & 0.064 & \pp &  1.315 $\pm$ \scriptsize 0.587 & 1.149 & \p &  0.108 $\pm$ \scriptsize 0.384 & \pp & 0.05 \\
			Mmorph-1 & 3 & 0.031 $\pm$ \scriptsize 0.012 & 0.027 & \pp &  1.304 $\pm$ \scriptsize 0.610 & 1.136 & \scriptsize 2.76 &  0.073 $\pm$ \scriptsize 0.192 & \pp & 0.07 \\
			Mmorph-2 & 3 & \textbf{ 0.031} $\pm$ \scriptsize 0.014 & \textbf{0.025} & -- & \textbf{ 1.303} $\pm$ \scriptsize 0.610 & \textbf{1.130} & -- & \textbf{ 0.002} $\pm$ \scriptsize 0.012 & -- & 0.07 \\ 
			\bottomrule
	\end{tabular}}
\end{table}

\para{Evaluation} 
We assess registration accuracy using two metrics: root mean square error~(RMSE) and end point error~(EPE). 
RMSE measures the intensity match between the reference image and the warped moving image, while EPE quantifies the discrepancy between the estimated motion and the motion utilized to generate synthetic moving images. 
The diffeomorphism of the deformation field is assessed by calculating the percentage of negative determinants~(NegDet).

\para{Results} Table~\ref{tab:Quanti_2D} offers a comparative analysis of our proposed methods (Mmorph-1,2) against two alternative strategies: Direct Registration~(Direct Reg) and Composition. The Direct Reg method serves as a baseline, where we directly estimate the Lagrangian between the reference frame $(t = 0)$ and last timeframe $(t = 2)$, without utilizing the temporal information within. 
Its high RMSE and EPE values suggest that the predicted motion significantly deviates due to the tag-jumping issue caused by the periodicity in the images.
The Composition approach mitigates the tag-jumping issue by simply composing the Eulerian motion to obtain the Lagrangian motion. 
However, it is less robust on complicated motion and has a higher percentage of negative determinants. 
Our proposed ``momenta, shooting, and correction" framework (Mmorph-1,2) outperforms the others in terms of registration accuracy and a lower percentage of negative determinants.
Interestingly, Mmorph-2, which employs a more accurate approximation of the BCH formula, achieves similar registration accuracy as Mmorph-1 but significantly reduces the percentage of negative determinants in the field, all while maintaining competitive computation time. 
Figure~\ref{fig:Quali_2D} provides a qualitative comparison of the different strategies on small, moderate, and large motion.\footnote{See Appendix~\ref{app:quali_2d_app} for additional visual results.}

\subsection{Tongue 3D tMRI Data}

\begin{figure}[!tb]
	\centerline{\includegraphics[width = 0.9\linewidth]{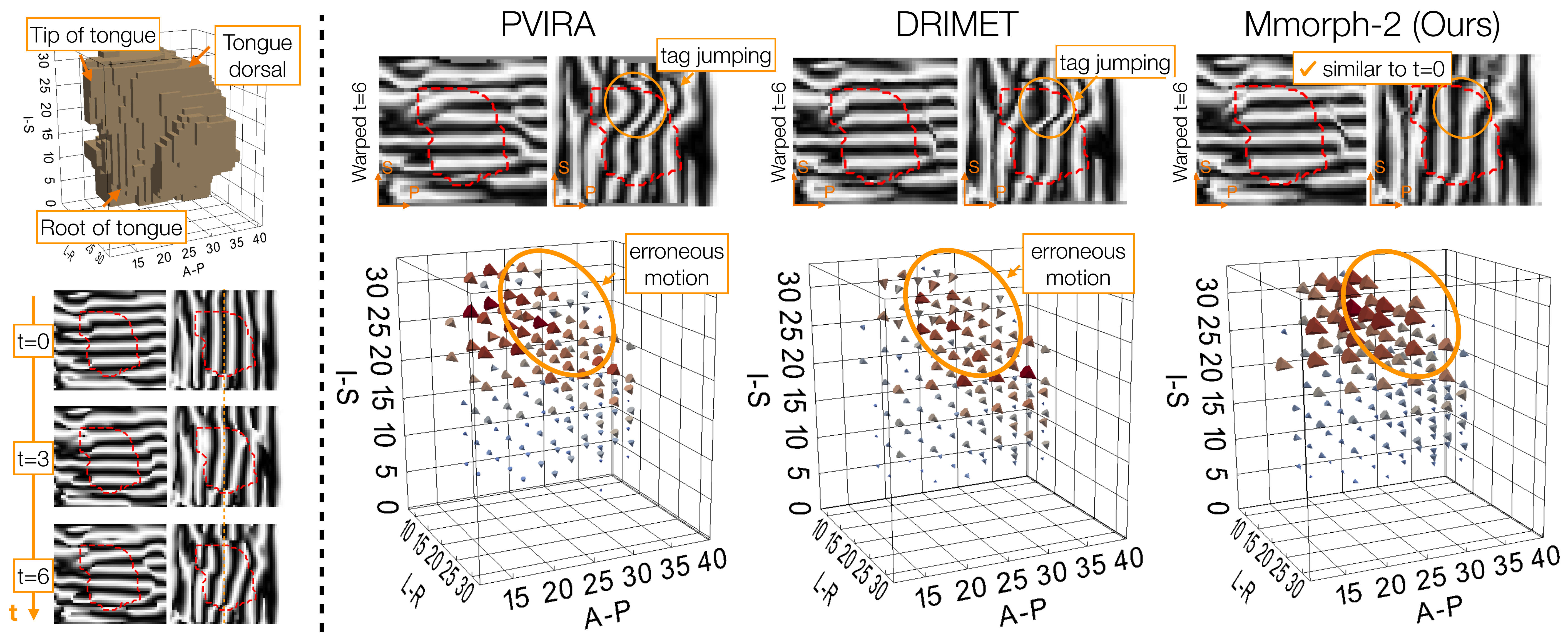}}
	\caption{Qualitative Results on 3D tMRI. The left panel depicts the 3D tongue shape and the sagittal slice with horizontal and vertical patterns. The right panel presents the results of three methods: PVIRA~\cite{PVIRA}, DRIMET~\cite{bian2023drimet}, and Mmorph-2 (our approach). The results of Mmorph-1 and the composition method have been omitted due to their visual similarity to Mmorph-2.}
	\label{fig:Quali_3D}
	%
\end{figure}

\begin{table}[!tb]
	\caption{Performance comparison on real 3D tMRI. The notation \(\mathbf{T}=\text{``a''}\) signifies that the method employs all neighboring pairs between the reference and moving frames for analysis. The p-values are adjusted by Bonferroni correction ($\times 5$).}
	\label{tab:Quanti_3D}
	\centering
	\resizebox{0.99\textwidth}{!}{%
		\begin{tabular}{cccccccccc}
			\toprule
			&  & \multicolumn{3}{c}{\textbf{RMSE} $\downarrow$ } & \multicolumn{3}{c}{\textbf{DetAUC} $\uparrow$} & \multicolumn{1}{c}{\textbf{NegDet (\%)} $\downarrow$} & \textbf{Time(s) $\downarrow$} \\ 
			\cmidrule(r){3-5} \cmidrule(lr){6-8} \cmidrule(l){9-9} \cmidrule(l){10-10}
			& $\mathbf{T}$ & Mean \scriptsize$\pm$ Std & Median & $p$ & Mean \scriptsize $\pm$ Std & Median & $p$ & Mean \scriptsize $\pm$ Std &  Mean \\ 
			\midrule
			PVIRA~\cite{PVIRA} & 2 & 0.211 \scriptsize $\pm$ 0.090 & 0.209 & \pp & \textbf{0.915} \scriptsize $\pm$ 0.030 & 0.917 & \scriptsize{0.55} & $2e^{-5}$ \scriptsize $\pm$ $6e^{-3}$  &  50.03 \\
			DRIMET~\cite{bian2023drimet} & 2 & 0.191 \scriptsize $\pm$ 0.075 & 0.185 & \p & 0.910 \scriptsize $\pm$ 0.058 & 0.919 & \p & $8e^{-6}$ \scriptsize $\pm$ $1e^{-7}$ & 0.15 \\
			\hdashline
			Composition & ``a'' & 0.192 \scriptsize $\pm$ 0.074 & 0.186 & \pp & 0.912 \scriptsize $\pm$ 0.057 & 0.923 & \scriptsize{3.5} & $3e^{-4}$ \scriptsize $\pm$ $1e^{-3}$ &  0.29 \\
			Composition & 3 & 0.190 \scriptsize $\pm$ 0.074 & 0.184 & \pp & 0.912 \scriptsize $\pm$ 0.055 & 0.923 & \scriptsize{4.45} & $5e^{-5}$ \scriptsize $\pm$ $2e^{-4}$ & 0.24 \\
			\hdashline
			Mmorph-1 & 3 & 0.188 \scriptsize $\pm$ 0.073 & 0.183 & \scriptsize{2.9} & 0.913 \scriptsize $\pm$ 0.053 & 0.920 & \scriptsize{2.95} & 0 \scriptsize $\pm$ 0 &  0.25 \\
			Mmorph-2 & 3 & \textbf{0.188 }\scriptsize $\pm$ 0.073 & \textbf{0.182}& -- & 0.913 \scriptsize $\pm$ 0.058 & \textbf{0.924} & -- & \textbf{0} \scriptsize $\pm$ 0 & 0.26 \\
			\bottomrule
	\end{tabular}}
\end{table}
\para{Materials} We further validate MomentaMorph using a real 3D MR dataset that includes 25 unique subject-phrase pairs. 
Participants spoke specific phrases during  tMRI scans, enabling tongue motion tracking. 
Each phrase lasts 1 second, during which 26 timeframes were acquired. 
The data was split subject-phrase-wise into training, validation, and test datasets in a 6:2:2 ratio.

\para{Training details} 
We applied a sinusoidal transformation~\cite{bian2023drimet} to the harmonic phase images~\cite{osman1999cardiac} as a preprocessing step. We adopted a similar hyperparameter tuning strategy as in our synthetic experiments. 
Notably,  $T$ timeframes were randomly selected from speech sequences during training. 
We find $T = 3$ effectively captures large tongue motion while preventing tag jumping.

\para{Evaluation} 
We do not employ end-point error~(EPE) due to the lack of true motion. 
We use the determinant of the area under the curve~(DetAUC)~\cite{bian2023drimet} to quantify the incompressibility. 
This measures how closely the Jacobian of the deformation approaches unity at each location. 
The motion between the reference frame (i.e., \(t=0\)) and all subsequent frames (i.e., \(t=1,2,\ldots,26\)) is estimated for evaluation. 
It is important to note that for multi-frame approaches (including ours and the composition approach) that utilize \(T \geq 3\) frames, additional \((T-2)\) intermediate frames are evenly sampled from the sequence.

\para{Results}
Figure~\ref{fig:Quali_3D} illustrates a case where the tongue initiates from a neutral position $(t = 0)$ and subsequently undergoes rapid backward and downward movements, as seen from the bent patterns at $(t = 6)$. 
Both PIVRA and DRIMET are challenged by substantial deformation, resulting in tag jumping. 
Interestingly, when compared between the composition strategies, composing fewer frames (T = 3) demonstrated a slight advantage over composing every neighboring frame (T = ``a''), which could be attributed to the drifting issue caused by successive interpolation. 
MomentaMorph achieves the best RMSE. 
In terms of incompressibility, MomentaMorph exhibits performance on par with the others. 
Consistent with the synthetic experiments, MomentaMorph had a lower percentage negative determinant than the composition strategy.

\section{Conclusion and Discussion}
In this work, we proposed a ``momentum, shooting, and correction'' framework that operates within the construct of Lie groups and Lie algebras, which provide a sound theoretical background, to estimate an accurate and biologically plausible Lagrangian motion field. 
The approach is validated through a 2D simulated dataset with elastic motion and a real 3D tMRI dataset. 
We believe that our approach is applicable beyond tMRI to general dynamic imaging. 
 As future work, we plan to validate the approach on cardiac imaging during heartbeat cycles and 4D lung CT scans during respiration, where substantial motion is often present.

%
%
%
 \bibliographystyle{splncs04}
 \bibliography{reference}

\begin{thebibliography}{10}
\providecommand{\url}[1]{\texttt{#1}}
\providecommand{\urlprefix}{URL }
\providecommand{\doi}[1]{https://doi.org/#1}

\bibitem{araslanov2021dense}
Araslanov, N., Schaub-Meyer, S., Roth, S.: Dense unsupervised learning for
  video segmentation. Advances in Neural Information Processing Systems
  \textbf{34},  25308--25319 (2021)

\bibitem{arsigny2006log}
Arsigny, V., Commowick, O., Pennec, X., Ayache, N.: A log-euclidean framework
  for statistics on diffeomorphisms. In: MICCAI. pp. 924--931. Springer (2006)

\bibitem{axel1989heart}
Axel, L., Dougherty, L.: {Heart wall motion: improved method of spatial
  modulation of magnetization for MR imaging}. Radiology  \textbf{172}(2),
  349--350 (1989)

\bibitem{axel1989mr}
Axel, L., Dougherty, L.: {MR imaging of motion with spatial modulation of
  magnetization}. Radiology  \textbf{171}(3),  841--845 (1989)

\bibitem{baker1897abel}
Baker, H.F.: Abel's Theorem and the Allied Theory: Including the Theory of the
  Theta Functions. University Press (1897)

\bibitem{voxelmorph2019}
Balakrishnan, G., Zhao, A., Sabuncu, M.R., Guttag, J., Dalca, A.V.: Voxelmorph:
  a learning framework for deformable medical image registration. IEEE TMI
  \textbf{38}(8),  1788--1800 (2019)

\bibitem{beg2005computing}
Beg, M.F., Miller, M.I., Trouv{\'e}, A., Younes, L.: Computing large
  deformation metric mappings via geodesic flows of diffeomorphisms. IJCV
  \textbf{61},  139--157 (2005)

\bibitem{bian2022learning}
Bian, Z., Jabri, A., Efros, A.A., Owens, A.: Learning pixel trajectories with
  multiscale contrastive random walks. In: CVPR. pp. 6508--6519 (2022)

\bibitem{bian2023drimet}
Bian, Z., Xing, F., Yu, J., Shao, M., Liu, Y., Carass, A., Zhuo, J., Woo, J.,
  Prince, J.L.: Drimet: Deep registration for 3d incompressible motion
  estimation in tagged-mri with application to the tongue. arXiv preprint
  arXiv:2301.07234  (2023)

\bibitem{bian2021deformable}
Bian, Z., Zhong, J., Hatt, C.R., Burris, N.S.: A deformable image registration
  based method to assess directionality of thoracic aortic aneurysm growth. In:
  Medical Imaging 2021: Image Processing. vol. 11596, pp. 724--731. SPIE (2021)

\bibitem{burris2022vascular}
Burris, N.S., Bian, Z., Dominic, J., Zhong, J., Houben, I.B., van Bakel, T.M.,
  Patel, H.J., Ross, B.D., Christensen, G.E., Hatt, C.R.: Vascular deformation
  mapping for ct surveillance of thoracic aortic aneurysm growth. Radiology
  \textbf{302}(1),  218--225 (2022)

\bibitem{campbell1926course}
Campbell, J.E.: A course of differential geometry. Clarendon Press (1926)

\bibitem{chen2023survey}
Chen, J., Liu, Y., Wei, S., Bian, Z., Subramanian, S., Carass, A., Prince,
  J.L., Du, Y.: A survey on deep learning in medical image registration: New
  technologies, uncertainty, evaluation metrics, and beyond. arXiv preprint
  arXiv:2307.15615  (2023)

\bibitem{christensen1996deformable}
Christensen, G.E., Rabbitt, R.D., Miller, M.I.: Deformable templates using
  large deformation kinematics. IEEE TMI  \textbf{5}(10),  1435--1447 (1996)

\bibitem{dalca2018unsupervised}
Dalca, A.V., Balakrishnan, G., Guttag, J., Sabuncu, M.R.: Unsupervised learning
  for fast probabilistic diffeomorphic registration. In: MICCAI. pp. 729--738.
  Springer (2018)

\bibitem{fu2020lungregnet}
Fu, Y., Lei, Y., Wang, T., Higgins, K., Bradley, J.D., Curran, W.J., Liu, T.,
  Yang, X.: {LungRegNet: an unsupervised deformable image registration method
  for 4D-CT lung}. Medical physics  \textbf{47}(4),  1763--1774 (2020)

\bibitem{gomez2020analysis}
Gomez, A.D., Stone, M.L., Woo, J., Xing, F., Prince, J.L.: Analysis of fiber
  strain in the human tongue during speech. Computer methods in Biomechanics
  and Biomedical Engineering  \textbf{23}(8),  312--322 (2020)

\bibitem{hall2013lie}
Hall, B.C., Hall, B.C.: Lie groups, Lie algebras, and representations. Springer
  (2013)

\bibitem{hering2021cnn}
Hering, A., H{\"a}ger, S., Moltz, J., Lessmann, N., Heldmann, S., van Ginneken,
  B.: {CNN-based lung CT registration with multiple anatomical constraints}.
  Medical Image Analysis  \textbf{72},  102139 (2021)

\bibitem{horn1981determining}
Horn, B.K., Schunck, B.G.: Determining optical flow. Artificial intelligence
  \textbf{17}(1-3),  185--203 (1981)

\bibitem{ibrahim2011myocardial}
Ibrahim, E.S.H.: Myocardial tagging by cardiovascular magnetic resonance:
  evolution of techniques--pulse sequences, analysis algorithms, and
  applications. Journal of Cardiovascular Magnetic Resonance  \textbf{13}(1),
  1--40 (2011)

\bibitem{jenkinson2003fast}
Jenkinson, M.: {Fast, automated, N-dimensional phase-unwrapping algorithm}.
  Mag. Reson. Med.  \textbf{49}(1),  193--197 (2003)

\bibitem{LKUnet2022}
Jia, X., Bartlett, J., Zhang, T., Lu, W., Qiu, Z., Duan, J.: {U-Net vs
  Transformer: Is U-Net Outdated in Medical Image Registration?} In:
  International Workshop on Machine Learning in Medical Imaging. pp. 151--160.
  Springer (2022)

\bibitem{jonschkowski2020matters}
Jonschkowski, R., Stone, A., Barron, J.T., Gordon, A., Konolige, K., Angelova,
  A.: What matters in unsupervised optical flow. In: Computer Vision--ECCV
  2020: 16th European Conference, Glasgow, UK, August 23--28, 2020,
  Proceedings, Part II 16. pp. 557--572. Springer (2020)

\bibitem{knutsen2014improved}
Knutsen, A.K., Magrath, E., McEntee, J.E., Xing, F., Prince, J.L., Bayly, P.V.,
  Butman, J.A., Pham, D.L.: {Improved measurement of brain deformation during
  mild head acceleration using a novel tagged MRI sequence}. Journal of
  Biomechanics  \textbf{47}(14),  3475--3481 (2014)

\bibitem{liu2010shortest}
Liu, X., Prince, J.L.: Shortest path refinement for motion estimation from
  tagged mr images. IEEE TMI  \textbf{29}(8),  1560--1572 (2010)

\bibitem{im2grid2022}
Liu, Y., Zuo, L., Han, S., Xue, Y., Prince, J.L., Carass, A.: Coordinate
  translator for learning deformable medical image registration. In:
  International Workshop on Multiscale Multimodal Medical Imaging. pp. 98--109.
  Springer (2022)

\bibitem{ilogdemons2011}
Mansi, T., Pennec, X., Sermesant, M., Delingette, H., Ayache, N.: {iLogDemons:
  A demons-based registration algorithm for tracking incompressible elastic
  biological tissues}. International Journal of Computer Vision
  \textbf{92}(1),  92--111 (2011)

\bibitem{osman1999cardiac}
Osman, N.F., Kerwin, W.S., McVeigh, E.R., Prince, J.L.: {Cardiac motion
  tracking using CINE harmonic phase (HARP) magnetic resonance imaging}. Mag.
  Reson. Med.  \textbf{42}(6),  1048--1060 (1999)

\bibitem{rohlfing2003volume}
Rohlfing, T., Maurer, C.R., Bluemke, D.A., Jacobs, M.A.: {Volume-preserving
  nonrigid registration of MR breast images using free-form deformation with an
  incompressibility constraint}. IEEE TMI  \textbf{22}(6),  730--741 (2003)

\bibitem{shao2023analysis}
Shao, M., Xing, F., Carass, A., Liang, X., Zhuo, J., Stone, M., Woo, J.,
  Prince, J.L.: Analysis of tongue muscle strain during speech from multimodal
  magnetic resonance imaging. Journal of Speech, Language, and Hearing Research
   \textbf{66}(2),  513--526 (2023)

\bibitem{spoorthi2018phasenet}
Spoorthi, G., Gorthi, S., Gorthi, R.K.S.S.: Phasenet: A deep convolutional
  neural network for two-dimensional phase unwrapping. IEEE Signal Processing
  Letters  \textbf{26}(1),  54--58 (2018)

\bibitem{stone2021smurf}
Stone, A., Maurer, D., Ayvaci, A., Angelova, A., Jonschkowski, R.: Smurf:
  Self-teaching multi-frame unsupervised raft with full-image warping. In:
  CVPR. pp. 3887--3896 (2021)

\bibitem{vialard2012diffeomorphic}
Vialard, F.X., Risser, L., Rueckert, D., Cotter, C.J.: Diffeomorphic 3d image
  registration via geodesic shooting using an efficient adjoint calculation.
  International Journal of Computer Vision  \textbf{97},  229--241 (2012)

\bibitem{wang2022deep}
Wang, K., Kemao, Q., Di, J., Zhao, J.: Deep learning spatial phase unwrapping:
  a comparative review. Advanced Photonics Nexus  \textbf{1}(1),  014001 (2022)

\bibitem{PVIRA}
Xing, F., Woo, J., Gomez, A.D., Pham, D.L., Bayly, P.V., Stone, M., Prince,
  J.L.: Phase vector incompressible registration algorithm for motion
  estimation from tagged magnetic resonance images. IEEE TMI  \textbf{36}(10),
  2116--2128 (2017)

\bibitem{ye2021deeptag}
Ye, M., Kanski, M., Yang, D., Chang, Q., Yan, Z., Huang, Q., Axel, L., Metaxas,
  D.: Deeptag: An unsupervised deep learning method for motion tracking on
  cardiac tagging magnetic resonance images. In: CVPR. pp. 7261--7271 (2021)

\bibitem{yu2020motion}
Yu, H., Chen, X., Shi, H., Chen, T., Huang, T.S., Sun, S.: Motion pyramid
  networks for accurate and efficient cardiac motion estimation. In: MICCAI.
  pp. 436--446. Springer (2020)

\bibitem{NewStartPoint2023}
Yu, J., Shao, M., Bian, Z., Liang, X., Zhuo, J., Stone, M., Prince, J.L.: {New
  starting point registration method for tagged MRI tongue motion estimation}.
  In: Medical Imaging 2023: Image Processing. SPIE (2023)

\bibitem{zhang2019lookahead}
Zhang, M., Lucas, J., Ba, J., Hinton, G.E.: Lookahead optimizer: k steps
  forward, 1 step back. Advances in neural information processing systems
  \textbf{32} (2019)

\end{thebibliography}

\appendix

\section{Material \& Training Details}
\label{app:exp}

\subsection{Synthetic Image Simulation}

We synthesized a dataset of 10,000 ``movies'' which undergo elastic deformation. The choice of elastic deformation over incompressible deformation for validation is motivated by the fact that elastic deformation is more general and prevalent. 
Specifically, we put a grid of evenly distributed control points across the image and randomly jitter these points, with magnitudes drawn uniformly from a predefined range. 
A series of deformation fields are generated by subdividing the offsets and interpolating between the control points with B-splines. 
The spacing between control points is set to twice of maximum displacement to ensure the generated deformation is free of folding. 
We ensure that the maximum deformation between consecutive frames does not exceed the theoretical critical point for tag-jumping~($P/2$). 
Those deformation fields are subsequently used to deform simulated sinusoidal images (both horizontal and vertical patterns), resulting in a time-series of deformed images.
The number of the sinusoidal pattern, denoted as $f$, is set to 10, which implies that the periodicity of the sinusoidal patterns is $P = 9.6$. 
Maximum possible deformation between neighboring frames is set to half of the period. 
The data samples were divided into training, validation, and test datasets in a 6:2:2 ratio. 
In assessing the effectiveness of various approaches, we omitted the boundary region (by 10\% along each axes) to mitigate the randomness of boundary effects and the influence of the padding strategy.

\subsection{tMRI Imaging Settings}

The study used a 3T Prisma MR scanner (Siemens Healthcare) and a 64-channel head/neck coil to obtain images of the tongue's motion during speech. Participants repeated a phrase while tagged MR images were acquired with a field of view of $240 \times 240$ mm$^2$, TE = 1.47 ms, and TR = 36 ms.  Each data set includes a stack of images covering the entire tongue and surrounding tissues. Multiple repetitions of the speech task were performed to collect tagged data. The speech task was timed to a metronome repeated every 2 seconds. 
The MRI images, with an in-plane resolution of 1.875 $\times$ 1.875 mm and a slice thickness of 6 mm, were obtained in both sagittal (vertical and horizontal tags) and axial (vertical tags only) orientations. 

\subsection{Training Details}

We normalize the images in the range $[0, 1]$. 
We do not employ any  augmentation and observe no signs of overfitting. 
To ensure a fair comparison across different approaches, we determine the optimal loss weights $\alpha$, and $\beta$ independently for each approach through a grid search.  The  $\gamma$ is empirically set to $0.5$ for all experiments. 

For experiment on 2D synthetic data, $\alpha=0.008$, $\beta=0$. For scaling and squaring, we used 7 as the number of iterations for all model variants. In all experiments, we used the Adam optimizer with a batch size of one and a fixed learning rate of $3 \times 10^{-4}$ throughout training. We trained using a batch size of 1 and trained until convergence at 200K steps without observing overfitting. Training time took around 10 hours using a single Quadro RTX 5000 GPU. 

For experiment on 3D tongue data, $\alpha=0.001$, $\beta=0.3$, the images are manually cropped to include the tongue region and then zero-padded to $64 \times 64 \times 64$. The (mean,std) of width, height, and depth of tongue region before zero-padding are: (48.3, 4.2), (56.5, 2.9), (54.2, 3.7), respectively. We used the Adam optimizer with a batch size of one and a fixed learning rate of $3 \times 10^{-4}$ throughout training. We trained using a batch size of 1 and trained until convergence at 500k steps without observing overfitting. Training time took around 48 hours using a single Quadro RTX 5000 GPU. 

\section{Additional Qualitative Results}
\label{app:quali_2d_app}
\begin{figure}[tb]
	\includegraphics[width= \linewidth]{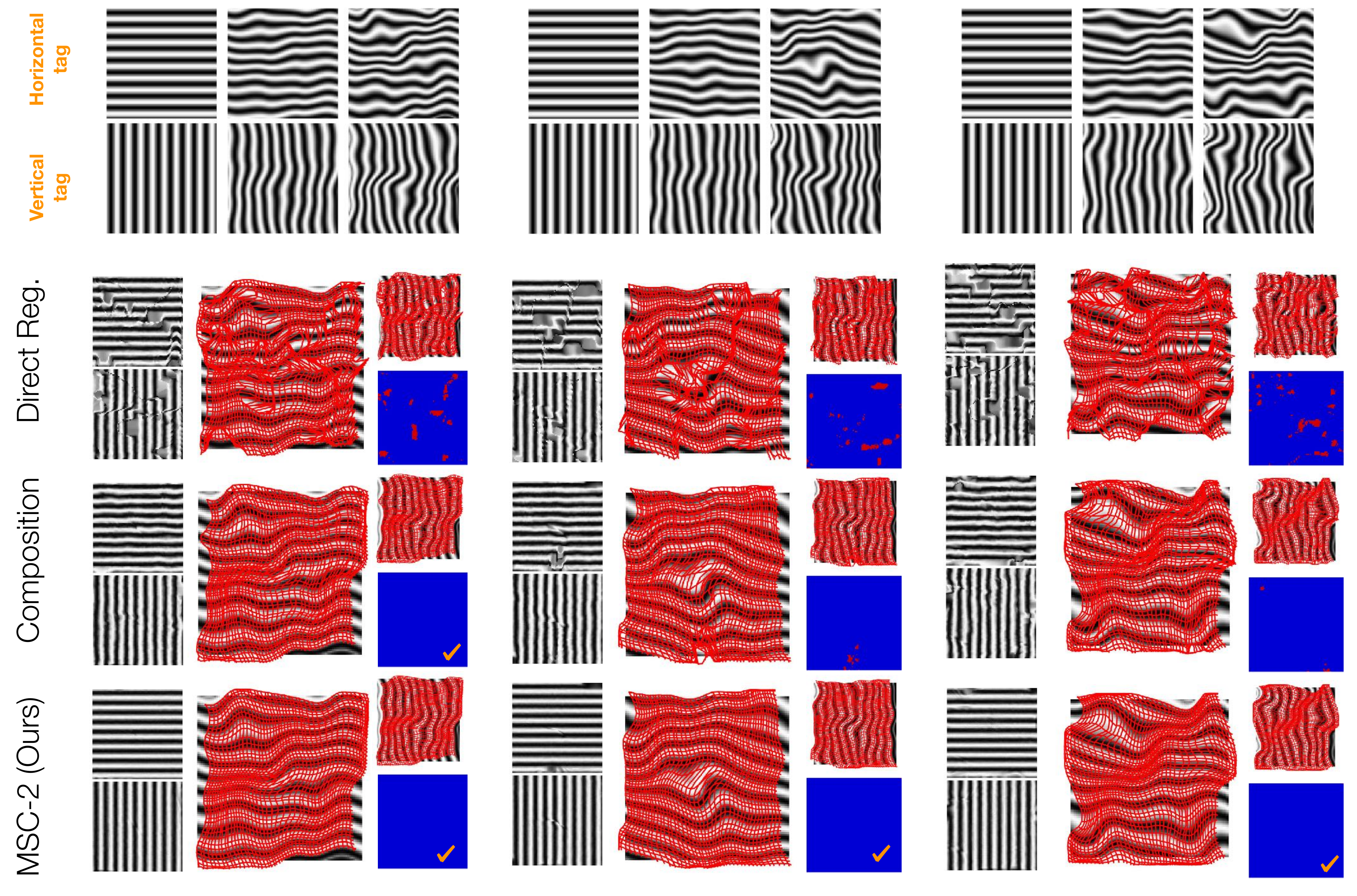}
	\caption{ Additional qualitative results on synthetic data. Row 1 indicates the input images sequence. The three rows below are the corersponding outputs from given the inputs. The checkmark $\checkmark$ on the Jacobian determinant map indicates diffeomorphism. }
	\label{fig:Quali_2D_app}
\end{figure}
Fig.~\ref{fig:Quali_2D_app} shows additional qualitative results.

\end{document}